\documentstyle[12pt]{article}
\begin{document}

 \baselineskip=0.85truecm

\title{Are Galactic Rotation Curves Really Flat?\footnote
{UCONN 96-04, May 1996}}

\author{\normalsize{Philip D. Mannheim} \\
\normalsize{Department of Physics,
University of Connecticut, Storrs, CT 06269} \\ 
\normalsize{mannheim@uconnvm.uconn.edu} \\}

\date{}

\maketitle

\begin{abstract}

In this paper we identify a new regularity in the systematics of 
galactic rotation curves, namely we find that at the last detected 
points in galaxies of widely varying luminosity, the centripetal 
acceleration is found to have the completely universal form 
$v^2/R=c^2(\gamma_0+\gamma^{*}N^{*})/2$ where $\gamma_0$ and 
$\gamma^{*}$ are new universal constants and $N^{*}$ is the amount 
of visible matter in each galaxy. This regularity points to a role 
for the linear potentials associated with conformal gravity, with the 
galaxy independent $\gamma_0$ term being found to be generated not 
from within individual galaxies at all, but rather to be of 
cosmological origin being due to the global Hubble flow of a 
necessarily spatially open Universe of 3-space scalar curvature 
$k=-(\gamma_0/2)^2=-2.3 \times 10^{-60}$cm$^{-2}$.

\end{abstract}


In discussions of the dynamics of galactic rotation curves it is usually 
assumed that rotation curves are asymptotically flat at large radial 
distances, and that whatever is responsible for this non-Keplerian behavior 
is itself just a purely local phenomenon which arises solely from within the 
galaxies themselves. In this paper we challenge these two widely accepted 
notions, and show that once the luminous Newtonian contribution is subtracted 
out, the resulting velocity discrepancies in individual galaxies are not 
merely actually growing (and quite rapidly in fact) with distance at the 
largest available radial distances, but, moreover, they are actually growing 
in a universal manner. Beyond being an interesting model independent 
phenomenological regularity in and of itself, and beyond being one which 
dark matter models of rotation curves should therefore be expected to 
account for, this regularity also points to a role for cosmology in the 
elucidation of rotation curves, as well as to the possible relevance of 
the conformal gravity theory of Weyl which is currently being explored by 
Mannheim and Kazanas as a candidate alternative to the standard dark matter 
paradigm. Moreover, with the apparent failure so far of the epochal 
gravitational microlensing observations to conclusively confirm the 
existence of the copious spherical dark matter halo that the Milky Way 
galaxy is widely believed to possess, the issue of alternate gravity even 
appears to have acquired some urgency. However, since the whole issue of
alternate gravity remains controversial at the present time, we shall 
begin by first looking for possible model independent clues in the data 
themselves.

In their analysis of available HI rotation curve data, Casertano and van 
Gorkom (1991) pointed out that the data basically fell into three broad 
categories: the rotation curves of low luminosity galaxies were found to 
generally be rising at the last detected points, those of intermediate to
high luminosity galaxies to be flat, and those of the highest luminosity 
galaxies to be (mildly) falling. Out of the available 28 galaxy set 
Begeman, Broeils, and Sanders (1991) identified a particularly reliable 
11 galaxy subset, and since these 11 galaxies range by more than 1000 in 
luminosity (see Table (1)) while clearly exhibiting the Casertano and van 
Gorkom trend (see Fig. (1)), the 11 galaxy subset should indeed be regarded 
as typical.\footnote{While none of the low luminosity galaxies
currently show any flat rotation curve region at all, there is a 
noticeable turnover in one of these galaxies, viz. DDO 154. However, 
since this is the most gas dominated galaxy in the entire sample, random 
gas pressures could be making a substantial contribution to motions in 
the turnover region. We shall thus ignore any possible ramifications of 
these last few points here, though clearly if this turnover proves to be 
a real trend which is then reproduced in other low luminosity galaxies, 
it would eventually have to be accounted for, not only here in fact but 
even in the standard dark matter theory and in the MOND  
theory both of which anticipate a flattening not a drop in the DDO 
154 rotation curve.} While the lack of flatness of the low luminosity 
galaxies is quite apparent, nonetheless the flatness (or near flatness)
of all the other galaxies is so striking that the rise in the low
luminosity galaxies has essentially been discounted by the general
community as being in any way suggestive of a trend, and it is 
generally assumed that these curves will eventually flatten off, with 
asymptotically flat rotation curves now being the standard paradigm. 
However, closer examination of the data reveals a possibly different 
outcome. Instead of looking at the actual rotation velocities, it is 
instructive to look at the velocity discrepancy, viz. the excess of the 
measured rotation velocity over the luminous Newtonian contribution,
a quantity whose overall normalization is essentially uniquely fixed
once and for all by the inner region rotation curve alone. Indeed, as we 
see from Fig. (1), this velocity discrepancy is itself far from flat, and
in fact is actually rising in each and every galaxy in the sample at 
the last detected data points. Since this discrepancy itself is usually 
explained by a spherical dark matter halo, we see that while such halos
may eventually lead to asymptotic flatness, their contributions in the
detected regions are necessarily still rising at the farthest points, with 
the flat total velocities that they produce in the bright galaxies actually
being achieved by carefully fine tuning the halo contribution galaxy by 
galaxy (through the use of two free parameters per halo and thus no less 
than 22 in total for our 11 galaxy sample) to rise at just the same rate as 
the luminous matter contribution is falling. Even for isothermal sphere 
halos asymptotia is thus still some way off, with the case for flat velocity 
discrepancies not yet being mandated by any available rotation curve data.

Beyond the issue of the shape of the velocity curves one can also ask if
there is any regularity in the magnitudes of the velocities. For the flat
rotation curve galaxies there is indeed such a regularity, viz. the 
Tully-Fisher law, a phenomenologically established universal relation 
between the luminosity and the fourth power of the velocity dispersion in 
the observed flat rotation curve region. Moreover, these same galaxies also 
appear to possess a second form of universality which was first noted
by Freeman, namely that the most prominent spiral galaxies all seem to 
have a common central surface brightness, $\Sigma_0^F$. (In passing we note 
that while there also exist low surface brightness galaxies with  
$\Sigma_0<\Sigma_0^F$, there do not appear to be any galaxies with 
$\Sigma_0>\Sigma_0^F$, thus making $\Sigma_0^F$ an empirical upper 
bound on galaxies). While the brighter galaxies thus possess a great 
deal of universality in addition to having flat rotation curves, this 
universality is not enjoyed by the non-flat low luminosity galaxies. Thus
it would be of interest to find a universality which also involves the
low luminosity ones as well. Given the suggestive fact that the velocity 
discrepancies are actually rising in all the galaxies, we thus evaluate
the centripetal acceleration at the last data point in each galaxy (except
for DDO 154 for which we use the last point before the turnover). As we can 
see from the fourth column in Table (1) the total $(v^2/c^2R)_{tot}$ is
remarkably universal, varying only by a factor of 5 or so over the sample
and certainly not by the factor of 1000 by which the luminosity varies in
the same sample. Even more interesting is the net value $(v^2/c^2R)_{net}$
obtained after the Newtonian contribution is extracted out. As we see from 
Table (1) this quantity only varies by a factor of 4. Moreover, we see a
small but clear trend with increasing mass in the centripetal acceleration.
And in fact, as will become more apparent below, we find that we can 
parameterize this net acceleration according to the two component relation 
$(v^2/c^2R)_{net}=(\gamma_0+\gamma^{*}N^{*})/2$ where the two universal 
constants $\gamma_0$ and $\gamma^{*}$ take numerical values $3.06\times 
10^{-30}$cm$^{-1}$ and $5.42\times 10^{-41}$cm$^{-1}$ respectively, and 
where $N^{*}$ is the total amount of stellar (and gaseous) material in solar 
mass units in each galaxy. (While the present author was drawn to this 
regularity via the conformal gravity study presented below, this regularity 
is an interesting one in and of itself which now serves as a new constraint 
on all theories of rotation curves.) As regards this regularity, it is 
important to realize that there is nothing in any way significant about the
actual magnitudes of the radial coordinates, $R$, of the last detected 
points in the 11 galaxies, since their locations are fixed purely by the 
instrumental limits of the various detectors used in measuring the various 
gas surface brightnesses and not fixed by any dynamics associated with the 
galaxies themselves. Thus the magnitude of each last measured radial 
$R$ (a quantity which varies from 8 kpc to 40 kpc or so over the sample) 
is essentially arbitrary for the galaxies, and yet once the Newtonian 
contribution is removed, $v^2/R$ can nonetheless still be universally 
parameterized. As far as we can see, the only obvious way that this could
in fact happen would be if $v^2$ were in fact growing universally with $R$
so that the magnitude of $v^2/R$  would not in fact depend on where the last 
detected points just happened to be located within galaxies. This pattern
is clearly not one that one would expect with flat rotation curves, or even
in fact think to look for in such a paradigm, and would instead seem to 
point to potentials which if anything are actually growing (linearly) with 
distance rather than falling in the familiar Newtonian manner.

Since our phenomenological analysis points to a role for linear potentials
in elucidating rotation curves, it is immediately suggested to consider
conformal gravity which contains such linear potentials to see whether it 
can account for the rotation curve phenomenology we have now identified. 
While conformal gravity dates back to Weyl and Eddington and to the early 
days of relativity, that it might enable us to dispense with dark matter 
was recognized only recently by Mannheim and Kazanas on finding
(Mannheim and Kazanas 1989; see also Riegert 1984) the exact metric outside 
of a star in the conformal theory, viz. $ds^2=B(r)c^2dt^2-dr^2/B(r)-
r^2d\Omega$ where $B(r)=1-2\beta^{*}c^2/r+\gamma^{*}c^2r$. Since this 
metric generalizes not only Newton but Schwarzschild also, it thus not 
only meets the classic solar system General Relativity tests, but it also 
provides for departures from Newton-Einstein on distances large enough  
that the linear potential term might first make itself manifest. 
Indeed, integrating the stellar potentials $V^{*}(r)=-\beta^{*}c^2/r+
\gamma^{*}c^2r/2$ over the visible galactic disk provides a luminous matter 
galactic potential (characterized by acceleration $v^2/R=g_{gal}^{lum}=
g_{\beta}^{lum}+g_{\gamma}^{lum}$) which nicely fits the shapes of the 
rotation curves of our 11 galaxy sample (Mannheim 1993, Mannheim and Kmetko 
1996, Carlson and Lowenstein 1996), but not their overall normalizations, 
since such a galactic disk would on its own only generate an asymptotic 
contribution $v^2/c^2R=\gamma^{*}N^{*}/2$ and thus lack the $N^{*}$ 
independent $\gamma_0/2$ term found above in our phenomenological 
analysis of centripetal accelerations.

Apart from the fact that the $\gamma^{*}N^{*}/2$ term arises from a 
non-Newtonian potential, it is otherwise a completely standard, local
non-relativistic term which arises from the local galactic matter
distribution and which scales as the total galactic luminosity. However, 
the additional $\gamma_0/2$ term we require is on a very different
footing since it is luminosity independent. Since, moreover, its
magnitude given above is of order the inverse Hubble radius, it would thus
appear to have to have a global, cosmological origin, with cosmology thus 
needing to provide galaxies with a second linear potential in addition
to the one that they themselves internally generate. Now, quite remarkably, 
it was noted by Mannheim and Kazanas in their original 1989 paper (where they 
found the generalized exterior Schwarzschild solution discussed above) that 
cosmology does precisely that. Specifically, they noted the kinematic fact 
that the general coordinate transformation
\begin{equation}
\rho= 4r / (2(1+\gamma_0 r)^{1/2} + 2 + \gamma_0 r)~~~~~,~~~~~
t = \int d\tau / R(\tau) 
\label{Eq. (1)}
\end{equation}
effects the metric transformation
\begin{equation}
(1+\gamma_0 r)c^2dt^2-{dr^2 \over (1+\gamma_0 r)}-r^2d\Omega\rightarrow 
{(1+\rho\gamma_0/4)^2 \over R^2(\tau)(1-\rho\gamma_0/4)^2}
\left(c^2 d\tau^2 - {R^2(\tau) (d\rho^2 + \rho^2 d\Omega)\over
(1-\rho^2\gamma_0^2/16)^2}
 \right) 
\label{Eq. (2)}
\end{equation}
to thus yield a metric which is conformal to a RW metric with scale 
factor $R(\tau)$ and (explicitly negative) 3-space scalar curvature
$k=-\gamma_0^2/4$.\footnote{In passing we note that in the 
cosmology discussed in Mannheim 1992, 1995b an open Universe with 
explicitly (very) negative $k$ was in fact realized, with such a Universe 
not suffering from the flatness problem found 
in the standard cosmology.} Now, and this is the key point, in a geometry 
which is both homogeneous and isotropic about all points, any observer can 
serve as the origin for the coordinate $\rho$; thus in his own local rest 
frame each observer is able to make the general coordinate transformation of 
Eq. (\ref{Eq. (1)}) involving his own particular $\rho$. Moreover, since the 
observer is also free in conformal gravity to make arbitrary conformal 
transformations as well, that observer will then be able to see the entire 
Hubble flow appear in his own local static coordinate system as a
universal linear potential with a universal acceleration $\gamma_0c^2/2$ 
coming from the spatial curvature of the Universe. Now in that specific 
static coordinate system any other Hubble flow observer would see something 
entirely different and not recognize anything that would look like a simple 
universal linear potential at all. Only in his own explicit rest frame would 
any other observer be able to recognize such a universal linear potential. 
Thus, while the transformations of Eqs. (\ref{Eq. (1)}) and (\ref{Eq. 
(2)}) would not be useful for describing the Hubble flow motions of the 
individual galaxies themselves, they appear to be ideally suited for 
describing the internal orbital motions of the stars and gas within each 
galaxy, since each internal motion can be discussed independently in each 
galaxy's own rest frame. Thus it would appear that in conformal gravity each 
observer sees the general Hubble flow metric as a local universal linear
potential with a strength fixed by the scalar curvature of the Universe (a 
nicely time independent quantity unlike the time dependent Hubble parameter 
itself), with the matter in each galaxy now acting as test particles which 
are being swept through the Hubble flow.\footnote{ We thus see a crucial 
difference between relativistic and non-relativistic reasoning. In strictly 
Newtonian physics the only effect of any background would be to put tidal 
forces on individual galaxies, forces that would not account for the 
rotational motions of stars and gas but only to a departure therefrom. 
Relativistically however, since the background produces an effect at 
the center of each galaxy, the background therefore contributes to the 
explicit rotational motions of the stars themselves, to thus yield a 
previously unappreciated but nonetheless quite general consequence of 
curvature.}

In order to now combine the local and global linear potentials we need to
embed each local galaxy into the Hubble flow and solve the gravitational
equations of motion in the presence of $T^{\mu \nu}_{local}+
T^{\mu \nu}_{global}$.\footnote{It is the very presence of 
$T^{\mu \nu}_{local}$ and its associated local geometry (viz. standard 
static Schwarzschild coordinates) which dictates the appropriate general 
coordinate transformation needed for Eq. (\ref{Eq. (1)}).} Given the fact 
that gravity is weak within galaxies, we shall as a first approximation 
simply add the local and global metrics given above to yield the total weak 
gravity acceleration $v^2/R=g_{tot}=g_{gal}^{lum}+\gamma_0c^2/2$ which can 
now be fitted to data. With $\gamma_0$ and $\gamma^{*}$ taking fixed 
numerical values (found to be $3.06 \times 10^{-30}$cm$^{-1}$ and $5.42 
\times 10^{-41}$cm$^{-1}$ respectively in the fitting), the fits reduce to 
just one free parameter per galaxy, viz. the standard optical disk mass to 
light ratio (or equivalently the total amount of stars and gas per galaxy, 
$N^{*}$, in solar mass units). Since, unlike dark matter theory, our theory 
is based on parameters with an absolute scale, it is thus very sensitive to 
distance determinations to galaxies. Consequently, we first calculate the 
total velocity predictions (the dotted curves) in Fig. (1) using the 
distances (listed in Table (1)) quoted by Begeman, Broeils and 
Sanders (1991) (this paper also gives complete data references). Then, 
again following Begeman, Broeils and Sanders, we allow 
for typical uncertainties in the adopted distances to give modest distance 
shifts of up to $\pm 15\%$ or so.\footnote{While larger shifts can actually 
improve the fits a little in some cases, we have not allowed for shifts of 
more than this except for NGC 1560 for which a distance estimate of 3.7 Mpc 
($+23\%$) has actually been reported in the literature.} With the indicated 
percentage shifts in adopted distance, with the fitted $M/L$ ratios 
listed in Table (1), and with $g_{gal}^{lum}$ calculated solely from the 
known luminous galactic matter (stars and gas), we then obtain the full 
curve fits of Fig. (1), with the dashed and dash-dotted curves showing the 
velocities that the Newtonian $g_{\beta}^{lum}$ and linear $g_{\gamma}^{lum}
+\gamma_0c^2/2$ terms would separately produce.\footnote{That cosmology 
might impact on rotation curves had already been suggested by Mannheim 
(1995a) in a paper where only the $\gamma_0$ term was considered in addition 
to $g_{\beta}^{lum}$. It is only with the inclusion of the local 
$g_{\gamma}^{lum}$ as well that the fits can be brought completely into line 
with the data.} No dark matter is assumed, and as we can see from the fits, 
none would appear to be needed. Despite the fact that our model is a highly 
constrained one with very few free parameters, it nonetheless appears to 
have captured the essence of the data (our fits have smoothed out some of 
the structure in the data since we treated the optical disks as single 
exponentials for simplicity), and phenomenologically our fitting 
would appear to be competitive with that of both the standard dark matter 
model (with its unsatisfactory plethora of free parameters) and the MOND 
(Milgrom 1983) alternative. Of course, beyond the question of fitting, 
unlike either dark matter or MOND, our theory is a fully 
motivated output to a fully covariant theory rather than being merely a 
phenomenologically motivated input, and for that reason alone it is already 
to be preferred over the other contenders. Moreover, if our theory is in fact 
correct, then it provides us with an actual measurement of the scalar 
curvature of the Universe, something which despite years of intensive work 
has yet to be achieved in the standard theory.  
     
In Table (1) we also list the values for the velocity discrepancy 
$(v^2/c^2R)_{net}$ at the last detected points as calculated at the shifted 
adopted distances by subtracting out the associated luminous Newtonian 
contribution. As we can see from Fig. (1), $(v^2/R)_{net}$ is 
indeed remarkably well fitted by $g_{\gamma}^{lum}+\gamma_0c^2/2 \sim 
(\gamma^{*}N^{*}+\gamma_0)c^2/2$ for each and every galaxy in our sample; 
and that even while the quantity $\gamma^{*}N^{*}/2$ does vary
enormously with luminosity over our sample (see Table (1)), nonetheless the 
$\gamma_0/2$ term overwhelms it in all but the largest galaxies, so that 
$(v^2/c^2R)_{net}$ only shows a mild (but nonetheless significant) dependence
on galactic mass. Given the values for $\gamma_0$ and $\gamma^{*}$ that we 
obtain from the fits, we see that these two terms would contribute the same 
amount for galaxies with $N^{*}_{crit}=5.65 \times 10^{10}$ stars which is 
indeed toward the high end of our sample.\footnote{In passing it is 
intriguing to note that with $\gamma^{*}$ being identifiable as the 
coefficient of the linear potential put out by a typical star such as the 
sun, and with $N^{*}_{crit}$ falling right in the range where the prominent 
galaxies are located, the asymptotic linear potential produced by a typical 
galaxy will be of the form $V_{\gamma}^{lum}(r)=
c^2\gamma^{*}N^{*}_{crit}r/2$, i.e. numerically of order 
$V_{\gamma}^{lum}(r)=c^2\gamma_{0}r/2$. Since each local galactic potential 
becomes of order one on distance scales of order $r=1/\gamma_{0}$, the 
cooperative effect of all of the galaxies in actually producing the Hubble 
flow in the first place can thus reasonably be expected to produce a Universe 
whose natural distance scale is in fact $1/\gamma_{0}$.} Since our theory is 
based on rising potentials, it is at first sight puzzling that it is able to 
(universally) fit the flat high luminosity rotation curves at all. To explain 
this intriguing aspect of our theory we recall that for an exponential disk 
spiral with surface brightness $\Sigma(R)=\Sigma_0$exp$(-R/R_0)$ the pure 
luminous Newtonian contribution causes the rotation curve to peak at around 
$2R_0$ with a normalization which depends on $\Sigma_0$. If we now 
universally match $\gamma_0$ to the Freeman limit value $\Sigma_0^F$, then 
in a Freeman limit galaxy with $N^{*}_{crit}$ stars (i.e. a galaxy whose 
entire linear term is then also universally normalized to $\Sigma_0^F$), 
the value of the velocity at, say, $10R_0$ or so (a region where the linear 
term is already dominant) will be equal to its value at the $2R_0$ Newtonian 
peak. Further, since at around $6R_0$ the Newtonian contribution has dropped 
to about half its peak value while the linear contribution there is at about 
half of its value at $10R_0$, we thus get a total velocity at $6R_0$ equal 
in magnitude to its values at both $2R_0$ and $10R_0$, and thus a flat 
rotation curve from $2R_0$ all the way out to about $10R_0$. Freeman limit, 
$N^{*}_{crit}$ galaxies thus naturally balance the falling Newtonian 
contribution against the rising linear one and allow flatness to obtain 
out to around $10R_0$ or so before the ultimate rise required of the 
linear potential finally sets in. Further, since we have tuned $\gamma_0$ 
to $\Sigma_0^F$ in the same galaxies,\footnote{Given this correlation, it is 
plausible that the Freeman limit itself may ultimately arise as an upper 
bound on the galaxies which are generatable as fluctuations out of the 
cosmological background, a background which is indeed controlled 
by the $\gamma_0$ scale.} at around $10R_0$ the velocity obeys 
$v^4 \sim R_0^2(\gamma_0)^2 \sim R_0^2(\Sigma_0^F)^2 \sim \Sigma_0^F L$, 
which we recognize as the Tully-Fisher relation. The universal matching of 
$\Sigma_0^F$ to $\gamma_0$ thus leads to both flatness and Tully-Fisher in 
$N^{*}_{crit}$ galaxies.\footnote{At this juncture it is interesting to 
point out that it is possible to make some contact with Milgrom's 
MOND alternative. Specifically, for Freeman limit, $N^{*}_{crit}$ galaxies,
we note that in the region (near $6R_0$) where the total $\beta$ and total 
$\gamma$ terms are approximately equal (i.e. where $g_{\beta}^{lum} 
\sim (g_{\gamma}^{lum}+\gamma_0c^2/2) \sim \gamma_0c^2$), $g_{tot}$ takes 
the numerical value $2(\gamma_0c^2g_{\beta}^{lum})^{1/2}$, an expression 
which we recognize as being of the MOND form on identifying $4\gamma_0c^2$ 
(=$1.1\times 10^{-8}$cm sec$^{-2}$) with Milgrom's $a_0$ (a 
phenomenologically introduced parameter whose fitted numerical value is 
typically found to be $1.2\times 10^{-8}$cm sec$^{-2}$ ). With this
equivalence we see that while conformal gravity and MOND give very
different predictions in the region beyond $10R_0$, they nonetheless give
quite similar predictions in the region below $10R_0$ were most of the
current measurements have been made. From a theoretical viewpoint we note
that while Milgrom developed MOND in order to be able to use a universal 
acceleration to explain the universal Tully-Fisher relation, the particular 
$a_0$ dominated region form for MOND that he chose (viz. $g_{tot}=
(a_0g_{\beta}^{lum})^{1/2}$) was motivated by the additional assumption of 
asymptotically flat rotation curves (and thus asymptotic flatness even for 
the low luminosity rotation curves as well). As we now see, it is in fact 
also possible to use a universal acceleration to explain Tully-Fisher in a 
theory where flatness only obtains as an intermediate phenomenon, and not as 
an asymptotic one.} Moreover, recognizing the special status enjoyed by  
$\Sigma_0^F$ and $N^{*}_{crit}$ in our theory, we are now able to explain 
the trend found by Casertano and van Gorkom. Since the low luminosity 
galaxies are both sub Freeman and sub $N^{*}_{crit}$, the $\gamma_0$ term 
wins and the rotation curves start to rise immediately. (This parallels the 
trend identified in dark matter fits where the low luminosity galaxies are 
found to be overwhelmingly dark.) Since the intermediate galaxies are both 
close to Freeman and close to $N^{*}_{crit}$, their rotation curves are both 
very flat and Tully-Fisher. And since the highest luminosity galaxies have 
$N^{*}$ greater than $N^{*}_{crit}$, the $\gamma_0$ term is temporarily
overcome so that the curves actually display a mild initial fall (and the 
galaxies will still be close to Tully-Fisher unless $N^{*}$ is altogether 
larger than $N^{*}_{crit}$). In this regard a particularly interesting high 
luminosity case is NGC 2841 whose data go out to twice as many scale lengths 
as the other high luminosity galaxies. For it the rotation velocity is 
actually seen to peak in the inner region at 326 $\pm$ 3 km sec$^{-1}$, to 
drop to a low of 271 $\pm$ 2 km sec$^{-1}$ in the intermediate region and to 
then rise back to 294 $\pm$ 6 km sec$^{-1}$ at the largest distances, a 
behavior which is quite suggestive of the onset of a delayed 
rise.\footnote{NGC 2841 is also of 
interest in another regard, since it does not actually appear to obey the 
Tully-Fisher relation. (To be a Tully-Fisher galaxy NGC 2841 would have to be 
at an adopted distance of about 18 Mpc (Begeman, Broeils and Sanders 1991)
rather than at the Hubble law determined distance of 9.5 Mpc which we have 
used in this paper, and it is thus one of the few galaxies for which the 
Hubble law and Tully-Fisher distance determinations differ markedly.) Since 
NGC 2841 is the only galaxy in our sample for which $N^{*}$ is altogether 
larger than $N^{*}_{crit}$, it is thus the only high luminosity galaxy (or 
high rotation velocity - its velocities actually being altogether bigger than 
those of any of the other galaxies) in our sample which according to our 
theory should then not in fact be Tully-Fisher.} The conformal gravity theory 
would thus appear capable of explaining the general systematics of galactic 
rotation curves in a completely natural manner, and our study suggests that 
rising rather than flat rotation curves is actually the paradigm, with the 
luminous Newtonian contribution having inadvertently masked that fact in the 
higher luminosity galaxies. Moreover, through the cosmological connection we 
have presented, we believe we have made a case for the existence of a 
universal linear potential associated with the cosmological Hubble flow, an 
intriguing possibility which appears to eliminate the need for dark matter.
This work has been supported in part by the Department of 
Energy under grant No. DE-FG02-92ER40716.00.

\noindent
{\bf Figure Caption}

\medskip
\noindent
Figure (1). The predicted rotational velocity curves associated with 
conformal gravity for each of the 11 galaxies in the sample. In each graph 
the bars show the data points with their quoted errors, the full curve 
shows the overall (adopted distance adjusted) theoretical velocity 
prediction (in km sec$^{-1}$) as a function of distance from the center of 
each galaxy (in units of $R/R_0$ where each time $R_0$ is each galaxy's own 
optical disk scale length), while the dashed and dash-dotted 
curves show the velocities that the Newtonian and the linear potentials 
would separately produce. The dotted curves shows the total velocities that 
would be produced without any adopted distance modification. No dark matter
is assumed.

\medskip
\noindent
{\bf Table (1)}
\medskip
$$
\begin{array}{cccccccc}

 Galaxy &  Distance & Luminosity & (v^2/c^2R)_{tot} & Shift  &
(M/L) &  (v^2/c^2R)_{net}&
\gamma^{\star}N^{\star}/2 \\

 {}& (Mpc) & (10^9L_{B \odot}) & (10^{-30}cm^{-1})  &
(\%)& 
(M_{\odot}L_{B \odot}^{-1})& (10^{-30}cm^{-1})&
 (10^{-30}cm^{-1}) \\

{}&{}&{}&{}&{}&{}&{}&{} \\ 

 DDO\phantom{1}~154  &\phantom{0}3.80& \phantom{0}0.05 & 1.51& 
-11 & 0.71  & 1.49 \pm .04  &0.01 \\

 DDO\phantom{1}~170  &          12.01& \phantom{0}0.16 & 1.63&
-07 & 5.36  & 1.47 \pm .07  &0.04  \\

 NGC~1560&            \phantom{0}3.00& \phantom{0}0.35 & 2.70&
+23 & 2.01 & 1.68 \pm .13  &0.08  \\

 NGC~3109&            \phantom{0}1.70& \phantom{0}0.81 & 1.98&
    & 0.01  & 1.74 \pm .19  &0.03  \\

 UGC~2259&            \phantom{0}9.80& \phantom{0}1.02 & 3.85&
+15 & 3.62  & 1.99 \pm .26  &0.15  \\

 NGC~6503&            \phantom{0}5.94& \phantom{0}4.80 & 2.14&
    & 3.00  & 1.58 \pm .15  &0.46  \\

 NGC~2403&            \phantom{0}3.25& \phantom{0}7.90 & 3.31&
+15 & 1.76  & 2.04 \pm .17  &0.66  \\

 NGC~3198&            \phantom{0}9.36& \phantom{0}9.00 & 2.67& 
-15 & 4.78  & 2.23 \pm .13  &0.97  \\

 NGC~2903&            \phantom{0}6.40&           15.30 & 4.86&
+14 & 3.15  & 2.83 \pm .19  &1.80  \\

 NGC~7331&                      14.90&           54.00 & 5.51& 
-16 & 3.03  & 4.42 \pm .50  &3.39  \\

 NGC~2841&            \phantom{0}9.50&           20.50 & 7.25&   
    & 8.26  & 5.75 \pm .30  &4.76
 
\end{array}
$$

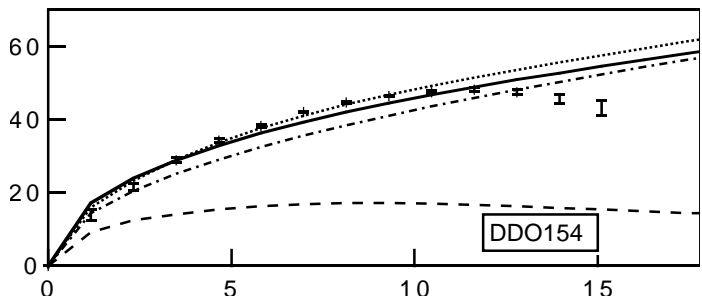
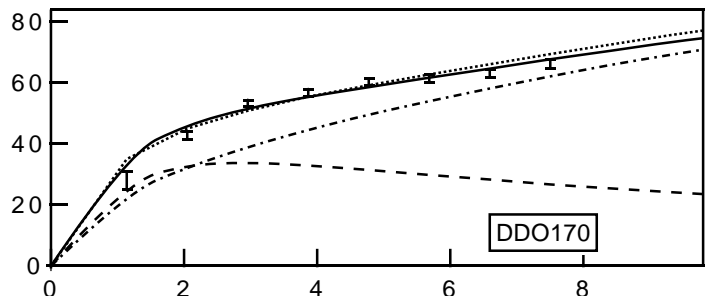
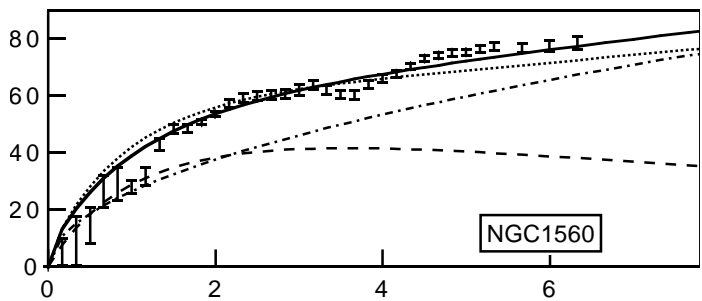
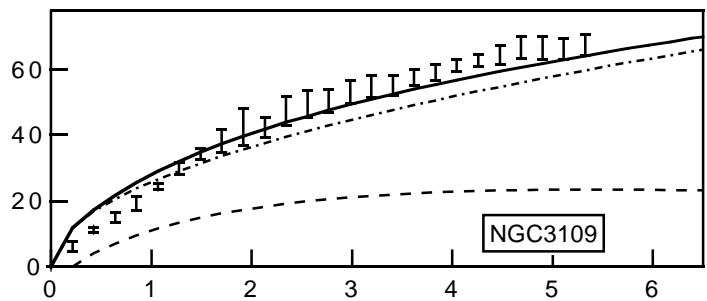
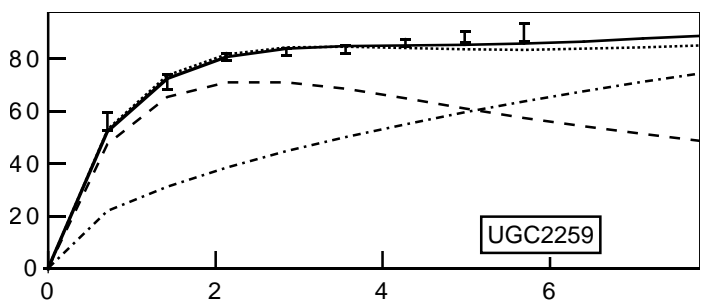
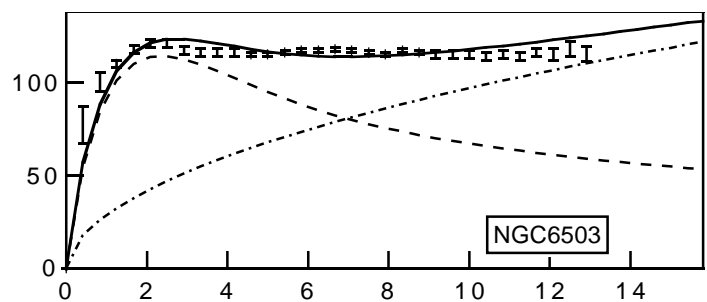
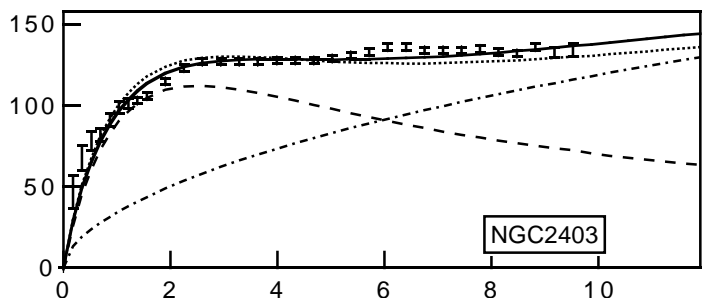
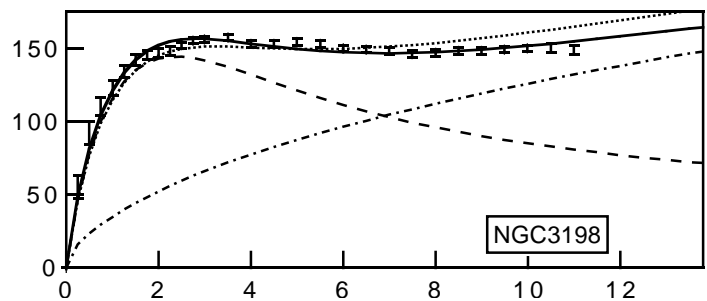
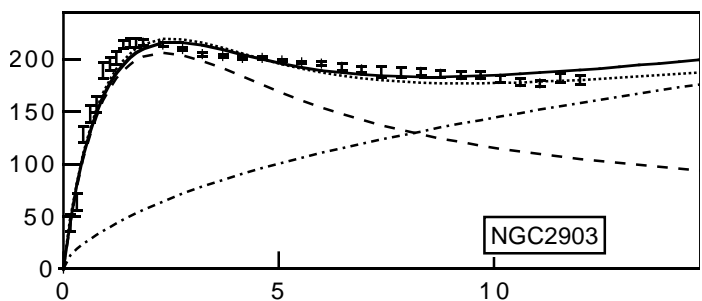
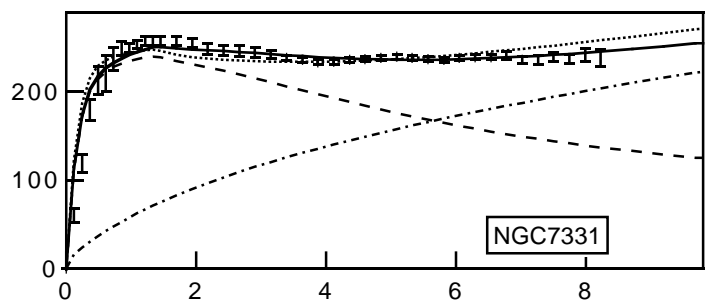
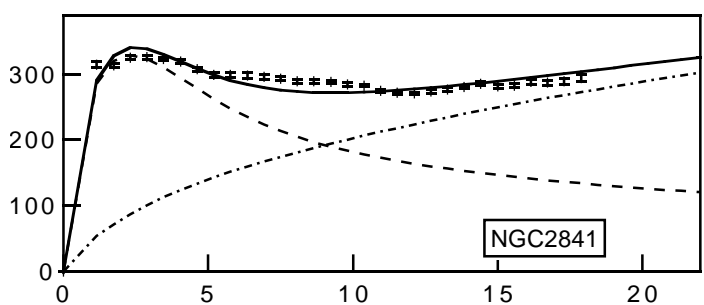
\end{document}